\documentclass{sig-alternate-05-2015}

\usepackage{algorithm}
\usepackage{algorithmic}
\usepackage{graphicx}
\usepackage{amssymb}
\usepackage{amsmath}
\usepackage{pifont}
\usepackage{multirow}
\usepackage{gensymb}
\usepackage[table]{xcolor}

\definecolor{Gray}{gray}{0.9}

\graphicspath{{./figures/}}

\begin{document}

\setcopyright{acmcopyright}



\conferenceinfo{ACM SIGKDD Workshop on Outlier Definition, Detection and Description on Demand (ODD 4.0)}{August 13-17, 2016, San Francisco, California, USA}


%

\title{Detection of Cyber-Physical Faults and Intrusions from Physical Correlations
}
%
%
%
%
%

\numberofauthors{5} 
%
\author{
%
%
\alignauthor
Andrey Y. Lokhov\\
       \affaddr{Center for Nonlinear Studies and Theoretical Division T-4}\\
       \affaddr{Los Alamos National Laboratory}\\
       \affaddr{Los Alamos, NM 87545}\\
       \email{lokhov@lanl.gov}
\alignauthor
Nathan Lemons\\
       \affaddr{Theoretical Division T-5}\\
       \affaddr{Los Alamos National Laboratory}\\
       \affaddr{Los Alamos, NM 87545}\\
       \email{nlemons@lanl.gov}
\alignauthor
Thomas C. McAndrew\\
	   \affaddr{Department of Mathematics and Statistics and Vermont Complex Systems Center}\\
       \affaddr{University of Vermont}\\
       \affaddr{Burlington, Vermont 05405}\\
       \email{thomas.mcandrew@uvm.edu}
\and  
\alignauthor
Aric Hagberg\\
       \affaddr{Theoretical Division T-5}\\
       \affaddr{Los Alamos National Laboratory}\\
       \affaddr{Los Alamos, NM 87545}\\
       \email{hagberg@lanl.gov}
\alignauthor
Scott Backhaus\\
       \affaddr{Materials Physics and Applications Division}\\
       \affaddr{Los Alamos National Laboratory}\\
       \affaddr{Los Alamos, NM 87545}\\
       \email{backhaus@lanl.gov}
}
\date{February 12, 2016}

\maketitle
\begin{abstract}
Cyber-physical systems are critical infrastructures that are crucial both to the reliable delivery of resources such as energy, and to the stable functioning of automatic and control architectures. These systems are composed of interdependent physical, control and communications networks described by disparate mathematical models creating scientific challenges that go well beyond the modeling and analysis of the individual networks. A key challenge in cyber-physical defense is a fast online detection and localization of faults and intrusions without prior knowledge of the failure type. We describe a set of techniques for the efficient identification of faults from correlations in physical signals, assuming only a minimal amount of available system information. The performance of our detection method is illustrated on data collected from a large building automation system. 
\end{abstract}

%
%


%
%

%
%



\vspace{1cm}

\section{Introduction}

Cyber-physical systems are physical networks, governed by the laws of physics, but regulated by a control system coupled to computer networks that transmit the information required to optimize and control the physical networks for reliability and efficiency \cite{sha2009cyber,shi2011survey}. Examples include, but are not limited to, smart grids, gas pipelines, civil infrastructures, autonomous automotive systems, automatic pilot avionics and process control systems. The interdependence of the cyber and physical networks makes the combined system more vulnerable to attacks; manipulation of the computer control network can leverage cyber-physical capabilities to cause damage or significantly degrade the performance of the critical infrastructure \cite{cardenas2008research,huang2009understanding}.

The ability to detect and localize failures or attacks represents an important step towards the design of resilient cyber-physical networks and strategies for implementation of certificates for proportional response. It is natural to expect that indications of intrusion or misbehavior in the cyber subsystem are present as anomalies in the physical network. This fact can be used for searching for outliers in the data streams collected by the sensors monitoring the state of the physical system -- a well-studied problem in a wide range of application domains \cite{6684530}. Although anomalous changes in individual signals can be an indication of a major failure or a crude attack, they do not capture more sophisticated scenarios of coordinated intrusions. Therefore, it is important to take into account information from the spatiotemporal correlations of anomalies of individual signals. Moreover, exploiting these correlations might enable probabilistic localization of the intruder or failure within the network, and hence serve as a basis for building a proper response.

We study the problem of detection and localization of disturbances based on the analysis of spatiotemporal correlations between physical data streams.  Our goal is to develop efficient methods for the detection and localization of failures within the cyber-physical system without reference to a predefined attack vector.  Failure events can be very diverse, while attacks become more and more creative and sophisticated, so the detection methodologies cannot be based on scripted scenarios. In addition, detection methodologies which do not exploit prior knowledge of the topology of the physical network will have a broader range of application. Therefore, we deliberately do not incorporate any specific aspects of the physical system architecture in the algorithm design. Further desired requirements for detection and localization algorithms include scalability (the number of signals and time measurements can potentially be very large), generality (we assume that the signals are heterogeneous and of diverse nature), robustness (the signals can be noisy and incomplete) and low computational complexity (to allow deployment of the algorithm in a fast online fashion).


Cyber-physical intrusion detection and response methodologies will improve at much faster rates when the development and refinement is closely coupled with real-world experimentation that validates strengths and reveals weaknesses. The simplicity and generality of the detection algorithms are very important  since they will allow for deployment in different cyber-physical systems. In this paper, we test our techniques on specific real-world data from an automated HVAC system in a large building at Los Alamos National Laboratory (LANL). We are planning to deploy and experimentally validate these methods on several other cyber-physical systems of importance to LANL.

We present a general protocol for detection and localization of disturbance which meet most of the aforementioned requirements. First, we develop a simple procedure for constructing a special correlation matrix out of detrended heterogeneous signals, making some assumptions on the anomaly signature we would like to be able to capture. Then, we use the correlation matrix to solve three crucial tasks: i) detection of the anomaly using spectral methods; ii) localization of a subset of anomalous nodes within the system using low-rank approximations and biclustering methods; iii) finally, identification of the functional role of the inferred anomaly based on the sensor labels. We validate our framework on experimental real-world data collected from a building automation system at LANL.

\section{Time Series Analysis and Correlation Matrix Construction}
\label{sec:Time_Series}

We consider the problem involving data from $N$ physical sensors indexed by $V$. For each sensor $i\in V$ we are given a time series $X_i(t)$ collected at times $t\in T$.  The data $X_i(t)$ can be heterogeneous real or integer valued signals and provides a (partial) description of a system.  We assume that the spatial and temporal relationships between the sensors are unknown, but that we do have access to sensor labels.  We also assume that the fluctuations of each time series in the system around their mean behavior during normal operations are essentially independent.

Formally, we say that during normal operations the observations $X_i(t)$ can be modeled as
\begin{equation}\label{eq:model}
X_i(t) = Y_i(t)+N_i(t)+S_i(t),
\end{equation} 
where the $N_i(t)$ represent the uncorrelated random noise, $S_i(t)$ is a potential signal of attack or failure (correlated between sensors) which is absent during normal operations, and $Y_i(t)$, which we call the trace, describes the idealized operation of the system without noise. 
When the system is attacked or experiences a fault the affected parts of the system $U \subset V$ are expected to move away from the trace, $S_i(t) \ne 0$ for $i \in U$.
We are interested in those cases when the signal is nonzero for a significant subset of sensors.  It may occur that for each individual sensor the failure signal is not directly observable, but that it can be detected and becomes statistically significant when the subset of affected sensors are taken into account collectively. In these cases, the differences between the trace and the corresponding observations will become related. In other words, since the $S_i(t)$ values corresponding to a particular disturbance event are likely to be correlated, we expect that the correlation relations will become apparent in the detrended signals $X_i(t) - Y_i(t)$
if the signal (e.g. attack or failure) occurs at $t=\tau$ and lasts for $T$ time steps. Our goal is to construct a suitable correlation matrix out of these time series which will enable the detection and localization of the undesirable changes in system state. 

\subsection{Detrending the Signals}

Unfortunately, the traces $Y_i(t)$ are \emph{a priori} unknown. In some cases they can be learned from an ensemble of repeating operations under normal behavior, but here we assume that this data might be unavailable. Thus we approximate the traces with a running mean,
\begin{equation}
\bar{X}_i(t):=\frac{1}{\tau_{\text{av}}}\sum_{t'=t-\tau_{\text{av}}/2}^{t+\tau_{\text{av}}/2}X_i(t'),
\end{equation} 
centered at $t$. This is a reasonable assumption if the traces $Y_i$ are fairly smooth: in this case, $\hat{X}_i(t)$ are smoothed  using the points $X_i(t')$ for $t'=t-\tau_{\text{av}}/2$ to $t+\tau_{\text{av}}/2$. This will not be a good assumption if the system changes modes of operation or otherwise undergoes rapid changes within the interval $[t-\tau_{\text{av}}/2,t+\tau_{\text{av}}/2]$.  

Note that although the use of the centered running mean requires the knowledge of the signal in the future, it produces better results with respect to the approach where the trailing mean is employed.  (A centered rolling mean approximates the trace with a linear function, while the trailing mean approximates the trace with a constant function.)
At the same time, an online detection algorithm based on the centered mean will have a time-lag of $\tau_{\text{av}}/2$. There is hence a trade off between the quality of approximation and the speed of detection.

It seems intuitive that the choice of smaller $\tau_{\text{av}}$ would introduce a smaller time-lag, and thus would lead to better results. On the other hand, $\tau_{\text{av}}$ should be large enough to average out the small fluctuations caused by the terms $N_i(t)$. A similar argument implies that $\tau_{\text{av}}$ should be chosen to be close in size to the expected duration of an attack or fault signal one would like to be able to detect: if $\tau_{\text{av}}$ is much larger than this scale, the signal will be likely to be averaged out. In practice, there is often a range of reasonable choices for the length $\tau_{\text{av}}$ of the sliding window; one should choose the one which satisfies the requirements on a desired maximum time-lag of detection. 

\subsection{Construction of the Correlation Matrix}

\begin{figure}[htb]
\centering
\includegraphics[width=1.0\columnwidth]{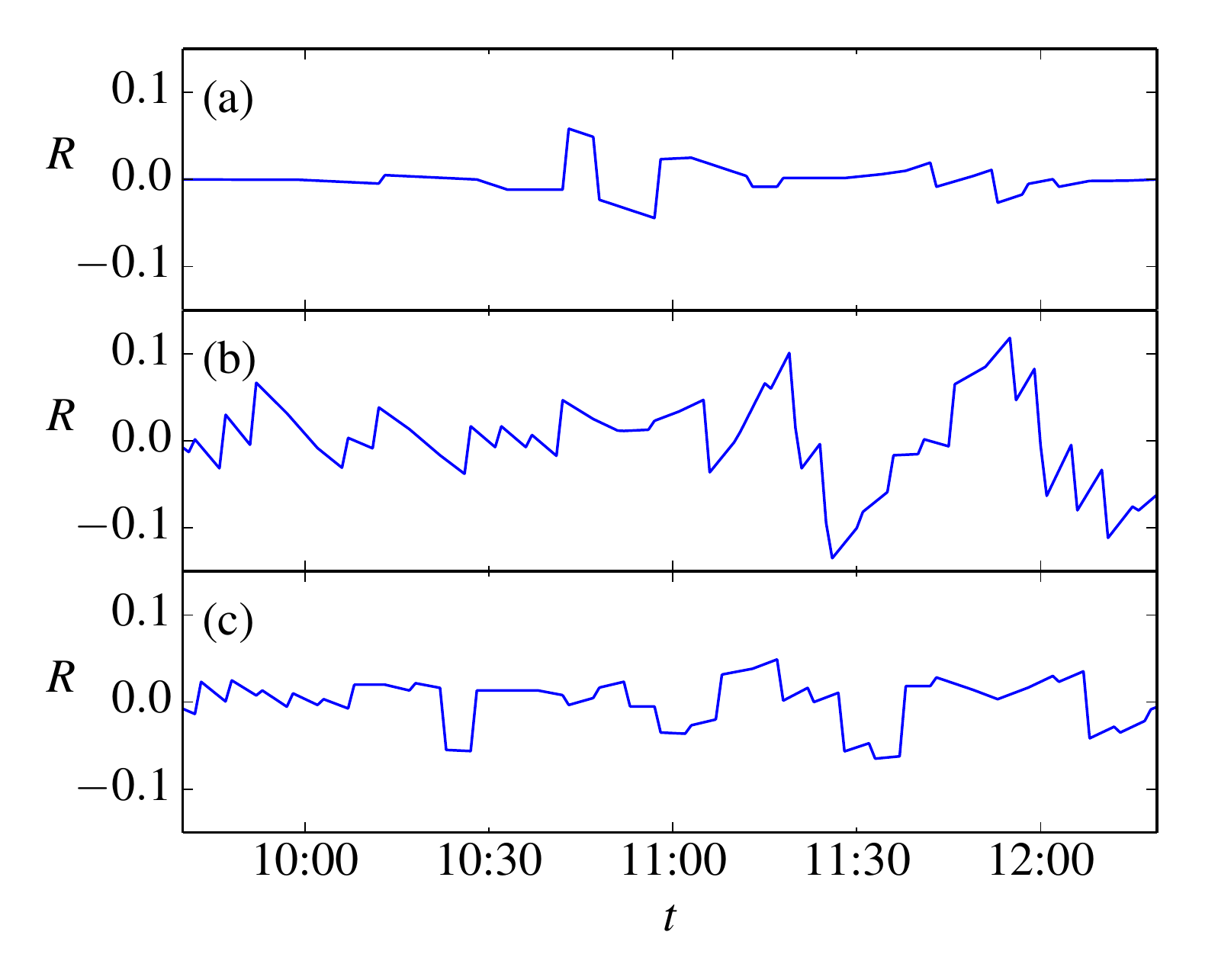}
\caption{Three residuals from a typical signal stream.  Signal (a) has $S(t)=0$ while signals (b) and (c) have correlated $S(t)\ne0$ due to an attack or failure.  The attack starts at approximatly 11:00 and some correlation can be observed between (b) and (c). 
The goal is to find and identify such correlated signals among the many recorded signals.
}
\label{fig:Error_Signals}
\end{figure}

We calculate correlation matrices from the residuals (an example is depicted in  Figure~\ref{fig:Error_Signals}) of the detrended data streams 
\begin{equation}
R_i(t) :=X_i(t) - \bar{X}_i(t).
\end{equation} 
At this point, one more parameter, the time interval $\tau_{\text{corr}}$ over which correlations are calculated, must be chosen. Ideally, this time window should be at least as large as the duration of the event we would like to detect. This time length, in general, is application dependent; typically, we are interested in the time scales which are a low multiple of $\tau_{\text{av}}$. Thus if the correlation window is determined to be of length $\tau_{\text{corr}}$, we calculate the Pearson correlation coefficient for each pair,
\begin{equation}
\xi_{ij}(t):= \frac{\sum\left(R_i(t')-\mu_{i,t}\right)\left(R_j(t')-\mu_{j,t}\right)}{\sqrt{\sum\left(R_i(t')-\mu_{i,t}\right)^2} \sqrt{\sum\left(R_j(t')-\mu_{j,t}\right)^2}},
\end{equation}
where each sum is taken from $t'=t-\tau_{\text{corr}}$ to $t$ and 
\begin{equation}
\mu_{i,t}:=\frac{1}{\tau_{\text{corr}}}\sum_{t'=t-\tau_{\text{corr}}}^t R_i(t).
\end{equation} 
This gives us the desired correlation matrix $M_{ij}(t)=\xi_{ij}(t)$ at each time instance. We are not interested in detecting the self-correlations which are trivially equal to one, so we put by definition $\xi_{ii}(t)=0$ $\forall i \in V$. 

With our setup under normal operations, when the data streams can be modeled as in Equation~\eqref{eq:model} with $S_i(t)=0$, we expect the detrended data streams to be uncorrelated,
\begin{equation}
\forall i\ne j,\;\;\mathbb{E}[\xi_{ij}(t)] = 0.
\end{equation} 
However, during an attack or failure we expect there to be a set of sensors $U\subset V$ such that  $S_i(t) \ne 0$ for $i \in U$, and hence
\begin{equation}
\forall r\ne s, \; r,s \in U, \;\; \mathbb{E}[\xi_{rs}(t)] = \sigma_{rs}>0,
\end{equation} 
since the non-zero signals $S_i(t), \; i \in U$ of the attack are supposed to have a similar behavior.

\section{Detection and Localization of Anomalous Submatrix}

In this section, we present a protocol for detecting and localizing a group of anomalously behaving devices within the physical network. Formulating the problem in the framework of submatrix localization, the detection step is done by monitoring the spectral gap in the correlation matrix spectrum. This method is universal and does not require any prior assumptions on the form of the noise and on particular normalization of the correlation matrix. We explore three approaches to the localization of the anomalous nodes: sparse PCA based on a low-rank approximation, and two biclustering methods for finding a submatrix with an elevated mean value.

\subsection{Detection of Anomalous Submatrix}
\label{sec:Detection}

Under normal conditions and low noise, the correlation matrix of the physical system might contain some structural information about the topology of the system. For instance, we can expect communities representing common functional roles or spatial locations of devices to have strong correlations. All other matrix elements should appear as noisy and uncorrelated values fluctuating around zero. When an anomaly occurs under the assumptions of Section 2 with a strong enough signal, one should witness the emergence of one single submatrix with a higher mean value.  As in the problem of detecting a single community in a graph \cite{Fortunato201075}, the change in the correlation matrix induced by the anomalous signal should be also visible in the spectrum of the correlation matrix.  In the ideal case, if the community is large enough, there is a spectral gap between the first and the second largest eigenvalues, and in addition, the principle eigenvector contains information about the location of the community.  We use the idealized case to gain intuition about the behavior of the real world system.

This intuition for the correlation matrices constructed from the real signals comes from rigorous analysis for ideal noise, which also illustrates the concept of a ``sufficiently strong signal'' used above. As an example, consider a rank-1 matrix with eigenvalue $\theta$, $P=\theta u u^{T}$, and suppose that we observe this matrix corrupted by a noise taking the form of a normalized $N\times N$ Gaussian Wigner matrix $W$, with zero-mean elements and variance of the off-diagonal elements equal to $1/N^{2}$. It is well known that the spectrum of $W$ converges to the semi-circle law with support $[-2,2]$. Let us denote the largest eigenvalue associated with the measurement matrix $P+W$ as $\lambda_{1}$, and the corresponding eigenvector as $u_{1}$. Depending on the ``signal strength'' $\theta$, the values of the largest eigenvalue and eigenvector of $P+W$ undergo a phase transition~\cite{benaych2011eigenvalues}. If $\theta > 1$, then in the large $N$ limit $\lambda_{1} \rightarrow 1 + 1/\theta$ is clearly separated from the bulk, and $\vert \langle u, u_{1} \rangle \vert \rightarrow 1 - 1/\theta^{2}$. In the opposite case $\theta \leq 1$, $\lambda_{1} \rightarrow 2$ and the associated eigenvector does not carry any useful information, being completely degraded by the noise, with $\vert \langle u, u_{1} \rangle \vert \rightarrow 0$. Similar results hold for the case of multiplicative noise.

\begin{figure}[ht]
\centering
\includegraphics[width=1.0\columnwidth]{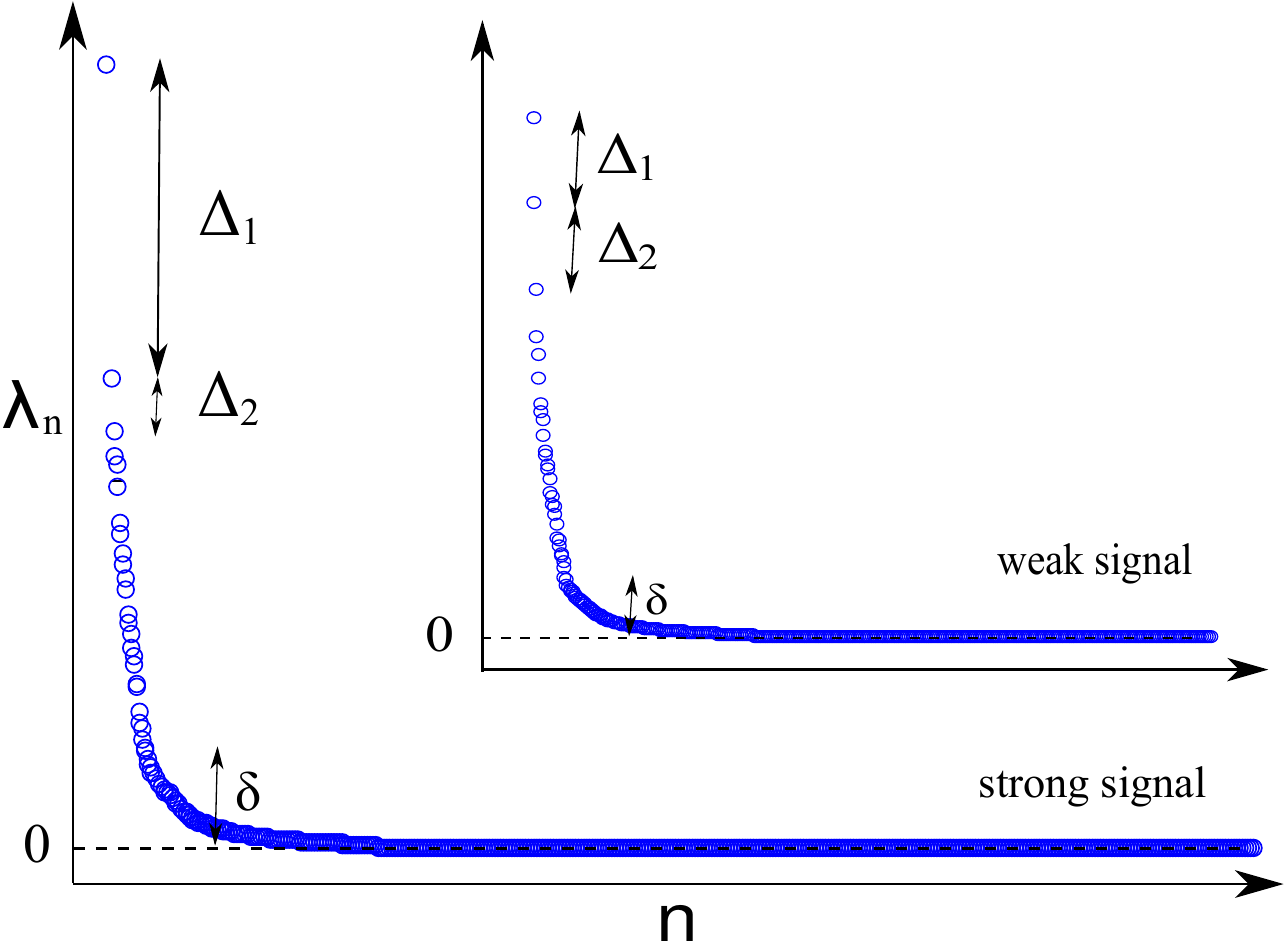}
\caption{A representation of a typical spectrum of a real-world correlation matrix in the presence of an anomaly (main figure) and with a weak anomalous signal (inset). In the first case, the condition \eqref{eq:Detection_condition} is satisfied, and hence we consider the outcome of the detection test as positive. In the case of weak signal, the level of noise does not allow us to conclude that an anomalous community of devices is present.}
\label{fig:Typical_spectrum}
\end{figure}

In a typical real-world situation, the spectrum of the correlation matrix in the presence of an anomalously correlated group of devices has a form presented in the main part of Figure~\ref{fig:Typical_spectrum}. There is a clear gap, separating two largest eigenvectors $\lambda_{1}$ and $\lambda_{2}$, and the nonzero values of eigenvalues $\lambda_{i}$ for $i \geq 2$, sorted by the order of magnitude, is entirely due to the noise. In the case of a weak signal, however, the picture can be similar to the inset of Figure~\ref{fig:Typical_spectrum}, where the presence of the spectral gap $\Delta_{1} = \lambda_{1} - \lambda_{2}$ does not seem to be so obvious. 

The important question is how to decide whether the gap is statistically significant. The challenge here is that we do not assume any prior information on the statistics of the trace and on the noise distribution; this setting has not been well studied in the literature so far. To address this question, we suggest the following detection criterion. Let $\Delta_{i} = \lambda_{i}-\lambda_{i+1}$ be the collection of spacings between successive eigenvalues of the correlation matrix. Following the assumption that the nonzero values of all eigenvalues but the largest one are entirely due to a random noise, we can empirically estimate the corresponding characteristic noise scale as 
\begin{equation}
\delta = \sqrt{\frac{1}{N-2} \sum_{1 < i < N} \Delta_{i}^{2}}.
\end{equation} 
Now our proposed detection certificate is as follows: we consider that the first eigenvalue is statistically well separated if
\begin{equation}
\Delta_{1} > \Delta_{2} + \delta.
\label{eq:Detection_condition}
\end{equation}
We count the opposite case as an absence of detection. The validity of this detection criterion will be checked in the Section \ref{sec:Experiments} involving real data examples.

\subsection{Localization Using the Low-rank Approximation}
\label{sec:Low-rank}

Once the detection certificate presented in Subsection~\ref{sec:Detection} yields a positive result, the next step is to localize the anomalously correlated elements of the system. The $K$ communities detection problem is often addressed using the low rank approximation \cite{coja2010graph}. In our case, a significant spectral gap $\Delta_{1}$ indicates that the hidden matrix can be localized by looking at the best rank 1 approximation $\widehat{M}$ of the initial matrix $M$,
\begin{equation}
\widehat{M} = \arg\min_{\widehat{M}} \Vert M - \widehat{M}\Vert_{F} \,\, \text{ s.t. } \text{rank}(\widehat{M})=1,
\end{equation}
where $\Vert \cdot\Vert_{F}$ is the Frobenius norm. The solution to this problem is well-known and is given by the singular value decomposition (SVD) of the matrix $M$, from which we retain only the leading singular value $\sigma$ and the corresponding singular vector $q$ \cite{eckart1936approximation}:
\begin{equation}
\widehat{M} = \sigma q q^{T}.
\end{equation}
Unfortunately, in general the resulting vector $q$ is not sparse, which does not allow us to identify the location of the anomalous nodes. Ideally, for detecting a group containing $k$ anomalous nodes, we would like to obtain a vector with only $k$ nonzero components, indicating their positions; this problem is often referred to as sparse PCA \cite{d2007direct}. While under a general low-rank assumption this problem is NP-hard, for the special case of rank 1 it can be solved analytically simply by sorting the elements of $q$, and retaining only $k$ largest elements \cite{papailiopoulos2013sparse, zhang2002low}, resulting in a $k$-sparse vector that we denote as $q_{k}$. The constant in the expression for $\widehat{M}$ is then simply given by $\sigma_{k} = q_{k}^{T} M q_{k}$.

Another difficulty comes from the fact that \emph{a priori} we do not know the size of the anomalous module. Sometimes, in order to find the optimal value of $k$, the so-called elbow method can be used \cite{thorndike1953belongs}. The idea is fairly simple; find the minimal $k$ such that the quality of approximation $\varepsilon_{k} \equiv \|M - \sigma_{k} q_{k} q_{k}^{T}\|_{F}$ is not increased ``too much'' when we make a step from $k$ to $k+1$. More precisely, the optimal $k$ is given by the minimal $k$ such that
\begin{equation}
\varepsilon_{k} - \varepsilon_{k+1} < \epsilon,
\label{eq:elbow_condition}
\end{equation}
where $\epsilon$ is some small constant, and the only parameter of the algorithm. The total complexity of the method is dominated by the complexity of the SVD-decomposition and is $O(N^{3})$ in the most general case.

\begin{figure}[htb]
\centering
\includegraphics[width=1.0\columnwidth]{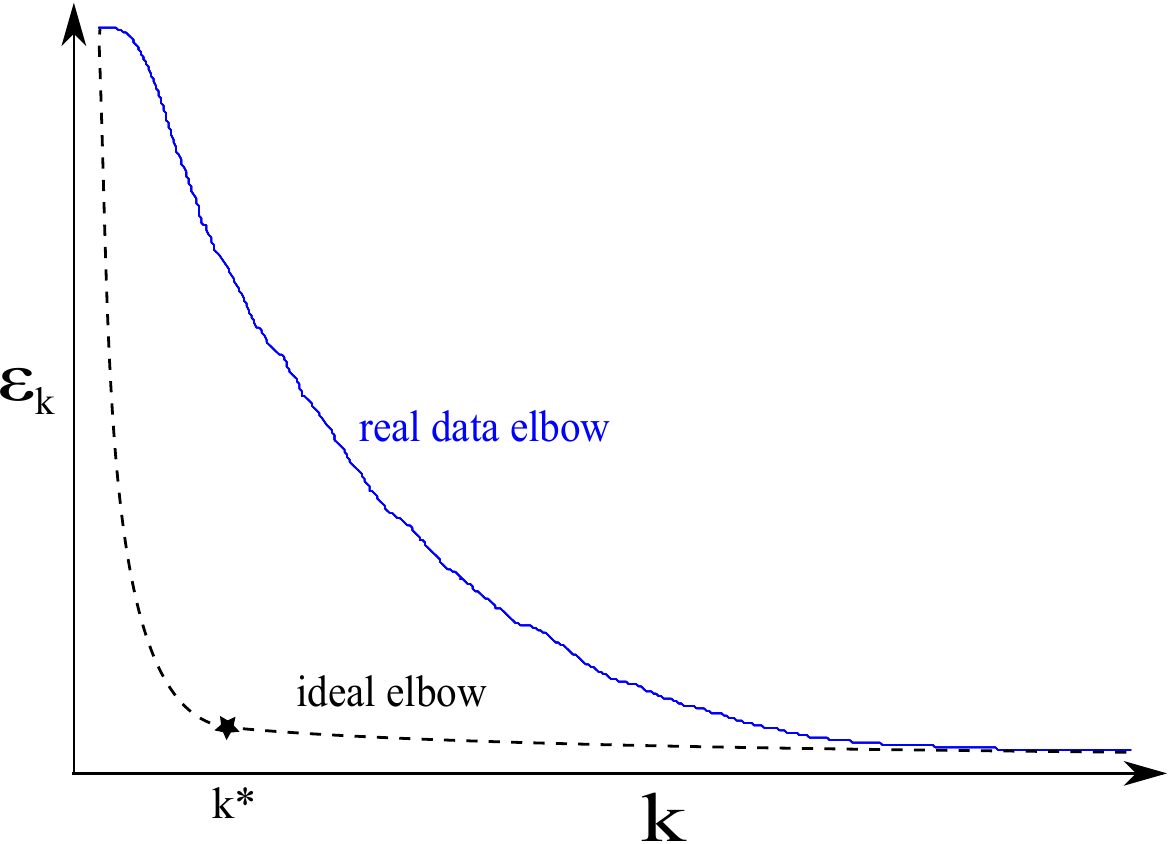}
\caption{An example of an ideal and real-world elbow diagram. In the case of relatively weak signals, the elbow plot produced from the real data does not have a pronounced cusp, which makes the identification of the optimal size of the group hard.}
\label{fig:Elbow_diagram}
\end{figure}

We expect the nonzero values of $q_{k}$ for the optimal $k$ to indicate the location of the nodes producing anomalous correlations. However, in the examples involving real data, the cusp on the elbow diagram might be not very pronounced in hard cases (see Figure~\ref{fig:Elbow_diagram} for an example), therefore, in practice it can be unclear how to select an appropriate $\epsilon$ and hence how to apply the condition \eqref{eq:elbow_condition}. At the same time it should be noted that at the end of the day we are not necessarily interested in inferring the whole set of anomalous nodes, but rather in understanding the cause of the anomaly. In this sense, one can choose to infer only a subset of anomalous sensors, but requiring a high level of confidence for this localization task; then the idea is to search for a subset of $k^{*}$ strongly correlated nodes. However $k^{*}$ can not be arbitrary small. Indeed, even in the idealized case there exist a practically achievable lower bound on the size of detectable community \cite{hajek2015information,deshpande2015finding} $k \gtrsim \sqrt{N}$. That is why the final suggested strategy consists in searching for a subset of most correlated sensors of size $k^{*} = \sqrt{N}$, and then in analyzing the corresponding group of devices using the tag data for determining the cause of the anomaly. This approach will be used in our experimental tests in Section \ref{sec:Experiments}, where an empirical evidence for the algorithmic failure in detection of communities of very small size will be presented.

\vspace{0.1cm}

\subsection{Localization via Biclustering Methods}
\label{sec:Biclustering}

In this part we discuss two efficient algorithms for localization of the anomalous subgraph of the physical network, which do not explicitly use the rank 1 assumption, but instead attempt to find a $k \times k$ submatrix with an elevated mean. The first one, called Large Average Submatrix ($\mathcal{LAS}$), has been introduced in \cite{shabalin2009finding} and analyzed in Ref.~\cite{bhamidi2012energy}, and consists in consecutive updates of $k$ rows and $k$ columns, starting from a random $k \times k$ submatrix and repeating the updates until a guaranteed convergence to a local maximum,  meaning that the resulting submatrix can not be improved by changing only its column or row set. A recently introduced improved version of this algorithm, analysed in \cite{gamarnik2016finding} and named Iterative Greedy Procedure ($\mathcal{IGP}$) follows a simple greedy scheme: starting by one randomly chosen row, we add the best columns and rows sequentially until a $k \times k$ submatrix is recovered. This algorithm outputs a provably better results, at least in the case of large Gaussian random matrices. In what follows, we test the performance of these algorithms on a real data set as a part of the localization procedure for finding the anomalously behaving group of nodes.

In order to get the best resulting submatrix, we use a multi-start procedure, initializing both algorithms $L$ times for given $k$, and retain the most significant submatrix. As before, the size of the hidden subgraph $k$ is unknown. In this case, again, we use $k^{*} = \sqrt{N}$ in order to find a smaller submatrix, representing the nodes which belong to the anomalous group of devices. The proposed method is summarized in Algorithm~\ref{alg:DetLoc}. The complexity of the Algorithm~\ref{alg:DetLoc} is dominated by the complexity of the localization step, and is equal to $O(N^{3})$ for the low-rank algorithm, to $O(I L N \ln N)$ for $\mathcal{LAS}$ and to $O(2 k^{*} L N \ln N)$ for $\mathcal{IGP}$, where $I$ is the number of iterations needed for convergence of the $\mathcal{LAS}$ scheme ($I \lesssim 1000$ for practical cases described here), and $L \gtrsim 10^{3}$ 
is the number of warm starts that we use in biclustering algorithms to achieve a desired precision of the best local maximum.

If the tag data (sensor labels) and/or additional topological information is available, one should be able to infer a possible cause of the failure by looking at the common factor uniting the selected nodes. In most cases, the selected basic devices are coupled to a single functional model or to a particular controller which might be at the origin of the fault and requires additional inspection.

\begin{algorithm}[tb]
   \caption{\textsc{Detection and localization of faults}}
   \label{alg:DetLoc}
\begin{algorithmic}
   \STATE {\bf Input:} $N$ time series $\{X_{i}\}_{i \in V}$, recorded in real time
   \STATE
   \STATE {\bf Correlation matrix:} compute $\{R_{i}(t)\}_{i \in V}$ and $M=\{\xi_{ij}(t)\}$ as described in Section \ref{sec:Time_Series}.
   \STATE {\bf Detection:} check for the condition \eqref{eq:Detection_condition} $\Delta_{1} > \Delta_{2} + \delta$.
   \STATE
   \IF{positive detection}
   \STATE {\bf Localization:} apply low-rank or biclustering algorithms on $M$, and infer a subset of $k^{*}$ anomalous nodes 
   \STATE {\bf Identification:} using the label data, infer the common cause of the failure
   \ENDIF
\end{algorithmic}
\end{algorithm}

\subsection{Tests with synthetic data}
\label{sec:Synthetic}

Prior to running tests on a real-world platform (next section), we examine the detection procedure on artificially-generated signals consisting of a mixture of correlated and uncorrelated one-dimensional random walks. In this idealized situation, we generate $N=900$ artificial signals as one-dimensional random walks starting from zero. We select  $k_{0}=50$ of them to be correlated and to represent an anomalous subgroup we would like to detect and identify. Uncorrelated random walks are lazy. With probability $p_{0}=0.9$, the  position at time $X_{i}(t+1)$ remains unchanged with respect to the previous time step $X_{i}(t)$, and with probability $p_{\pm}=0.05$ two positions separated by one time step satisfy $X_{i}(t+1) = X_{i}(t) \pm 1$. Correlated random walks are constructed as follows: they are related to one of the random walks (called the master random walk), at each time step independently repeating the step of the master random walk with probability $\rho = 0.5$, and otherwise behaving as an uncorrelated random walk.

\begin{figure}[htb]
\centering
\includegraphics[width=0.88\columnwidth]{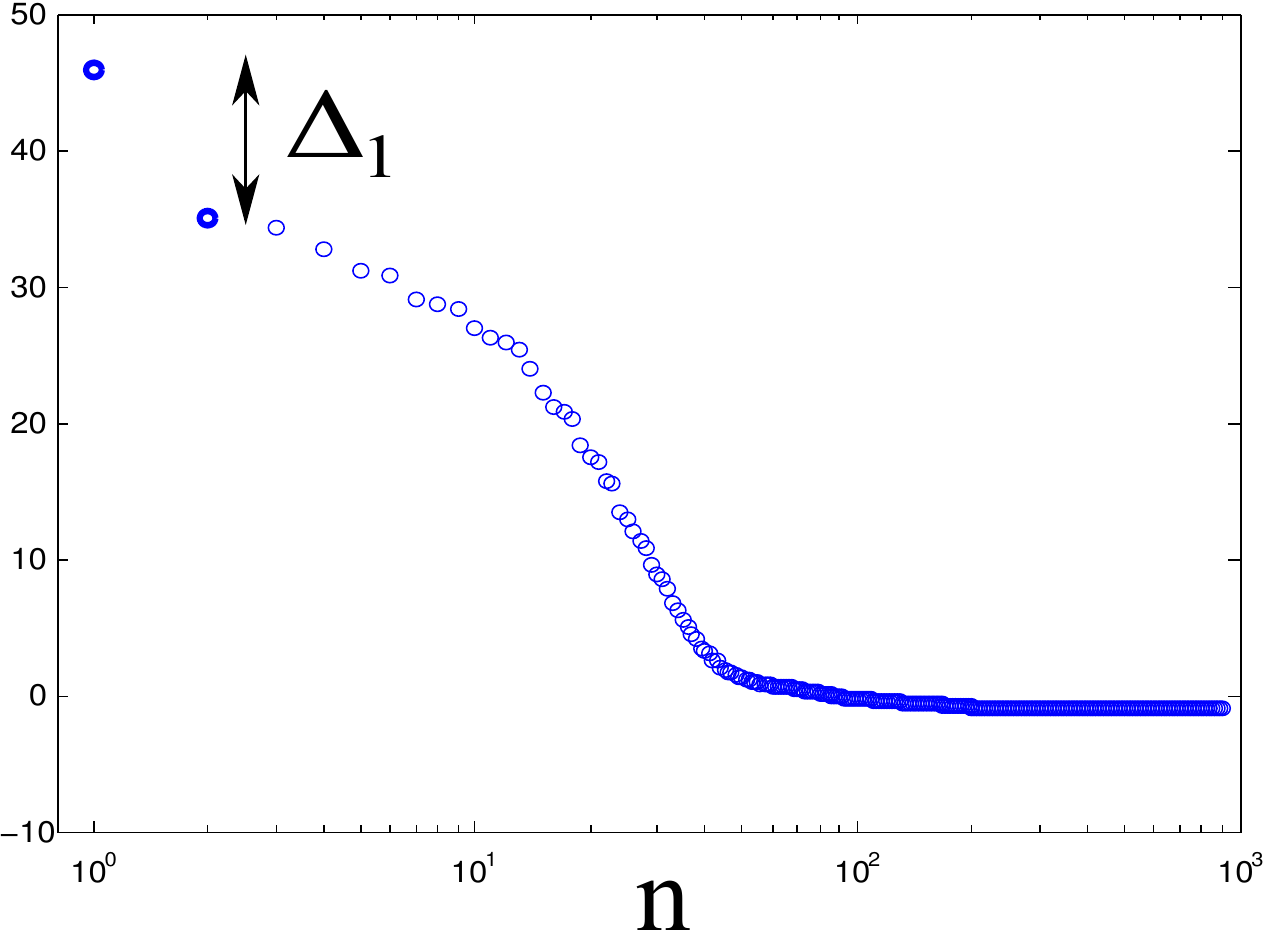}
\caption{The spectrum (on a semi-log scale) of the correlation matrix $M$ constructed from the total of $N=900$ artificially-generated signals, including $k_{0}=50$ correlated walks.  The correlated group of signals produces an identifiable gap $\Delta_1$ in the eigenvalue spectrum.}
\label{fig:Synthetic_signals}
\end{figure}

Let us now show the performance of Algorithm~\ref{alg:DetLoc} on this artificial signal ensemble. First, we detrend the data and construct the correlation matrix $M$ in the way described in Section~\ref{sec:Time_Series}; we choose $\tau_{\text{corr}}=200$, and the running mean is taken over the window $\tau_{\text{av}}=10$ time steps. The spectrum of $M$ is presented in Figure~\ref{fig:Synthetic_signals} and triggers a positive detection according to the criterion \eqref{eq:Detection_condition}. 

Next, we run the localization algorithms presented in Sections \ref{sec:Low-rank} and \ref{sec:Biclustering}. We find that for $k^{*}=\sqrt{N}=30$, all algorithms perfectly identify a subgroup of $30$ correlated signals. If we choose to search the correlated group with the (unknown) ground truth size $k_{0}=50$, then the low-rank approximation approach misidentifies $5$ signals, correctly counting the other $45$ as correlated. Both biclustering methods make only one mistake in this case; however, it requires a rather large number of warm starts ($L \simeq 3 \cdot 10^{4}$) in order to converge to the best solution, which makes the algorithm slightly slower compared to the SVD-based one. As we will see in the next section, the speed of convergence is a very important property for online deployment of the algorithm.

\section{Experiments with Real Data}
\label{sec:Experiments}

\subsection{System Description}

\begin{figure}[htb]
\centering
\includegraphics[width=1.0\columnwidth]{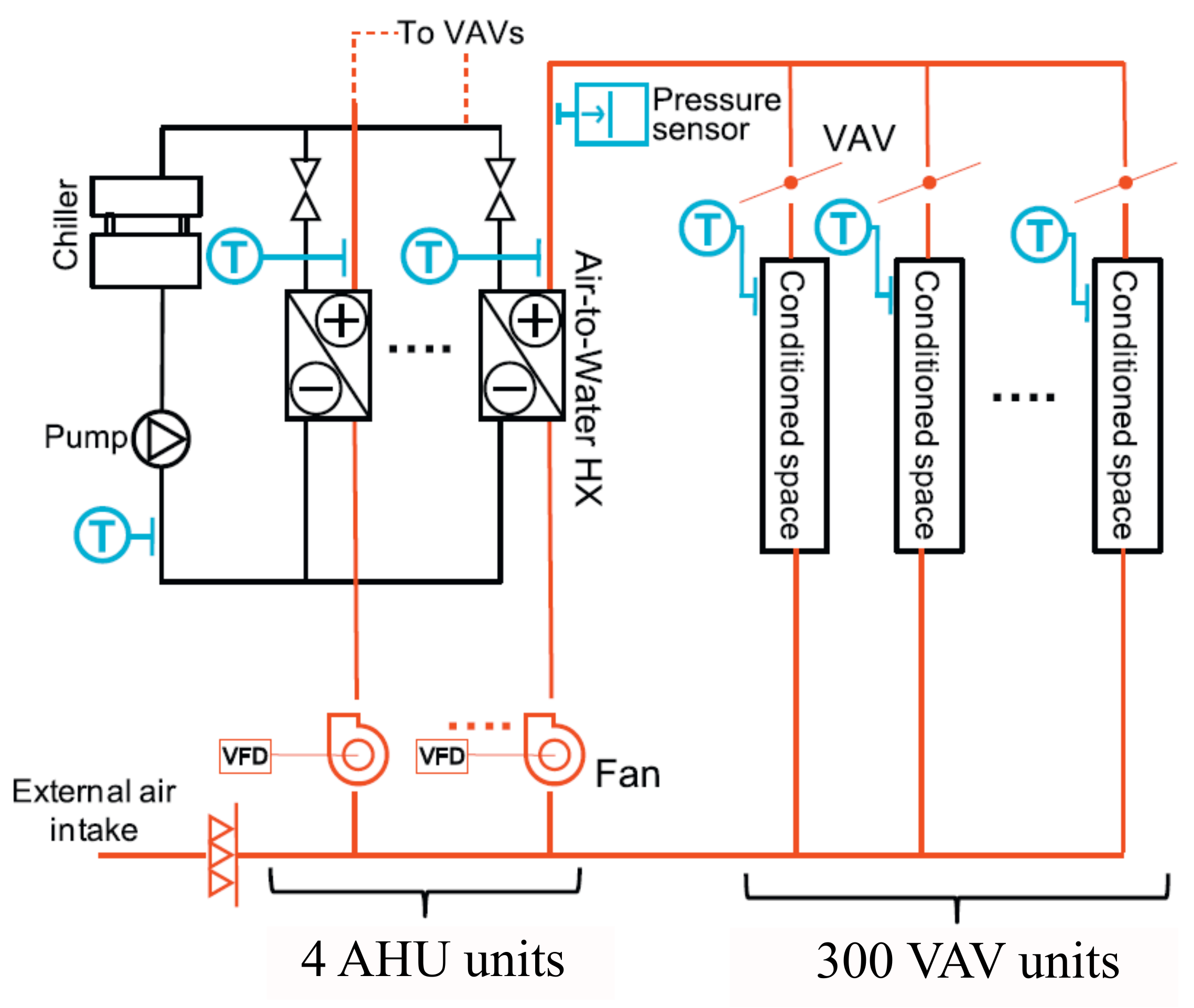}
\caption{A schematic representation of air conditioning (AC) system used in this work. The AC system includes two sets of loops: a water loop circulating water between the chiller and the air-to-water heat exchangers, and the air loops, where the fans in the air handling units (AHU) force the warm return air through the heat exchangers, and the cooled air is then delivered to the variable air volume (VAV) units. Thermostats (T) throughout the system provide input to the controllers that regulate the air flows supplied to the VAVs. The recorded temperature, airflow and valve opening position signals from all the sensors and fans are used as input data streams to our fault detection and localization algorithm.}  
\label{fig:Scheme}
\end{figure}

Large commercial air conditioning (AC) systems represent an attractive cyber-physical test case for fault detection and localization algorithms because they contain relatively sophisticated physical, control and communications architectures, and the available tag data can serve as a ground truth for discovered groups and modules. We collected and analyzed the data streams from the AC system in a 30 000 m$^2$ office building, with about 900 sensors located in the conditioned spaces. These sensors  record local temperature, airflow and valve opening positions. See Figure \ref{fig:Scheme} for a schematic representation of the system used in this study, which shares a common structure with a large number of commercial AC systems. A more in depth discussion of this AC layout is provided in the references \cite{beil2015round,goddard2014model}.  Altogether this constitutes a system of approximately $1000$ data heterogeneous data streams, sampled once per minute.

\begin{figure}[htb]
\centering
\includegraphics[width=1.0\columnwidth]{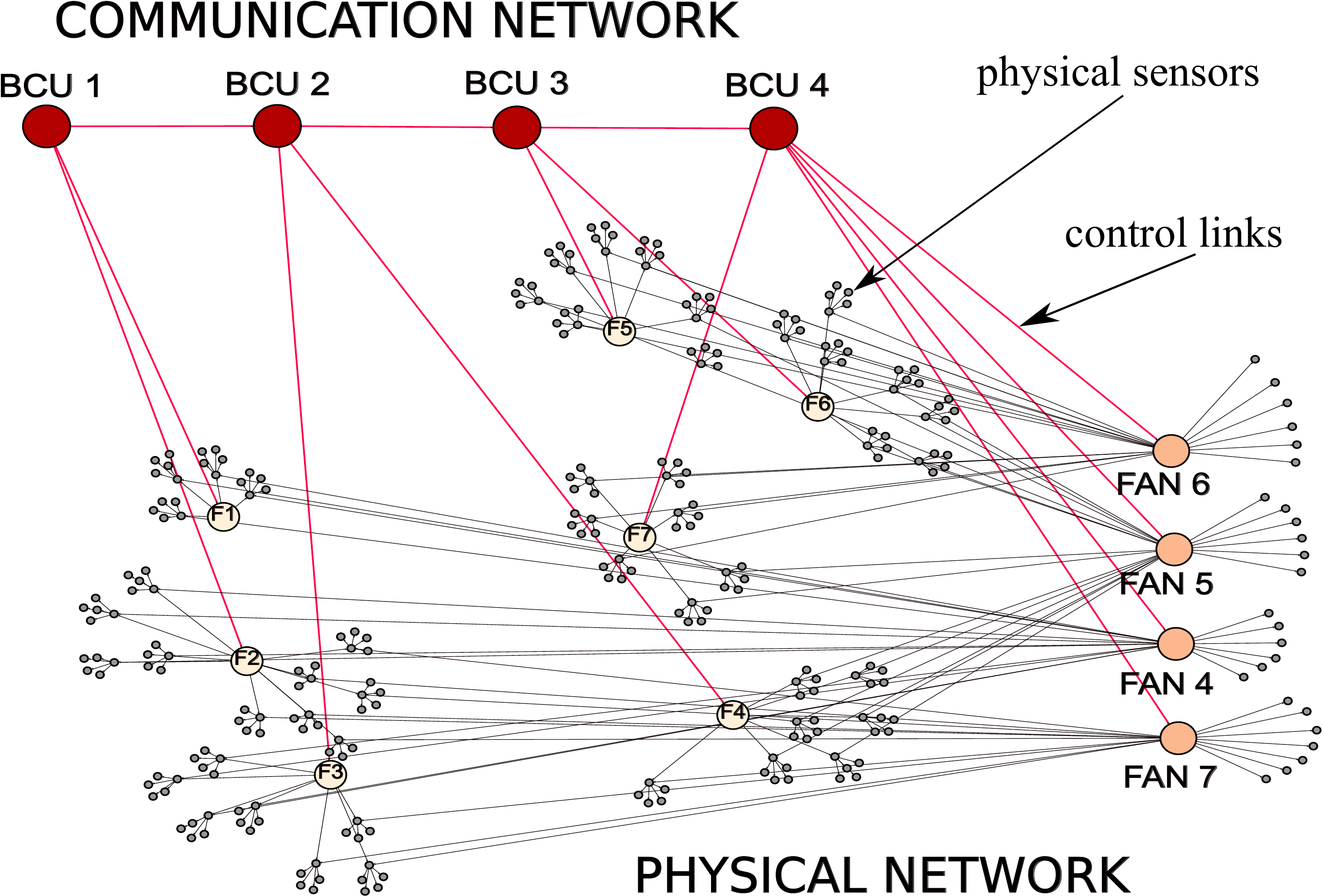}
\caption{Network representation of a part of the cyber-physical system considered in this work. The network reflects the spatial organization of the conditioned spaces, and includes a part of both physical and control links. Fan 6 is the anomalously behaving unit of the system.}
\label{fig:Tag_network}
\end{figure}

The variable air volume (VAV) units represent the air inlets to the cooled spaces, containing valves that regulate the chilled air flowing to the conditioned space. Different VAVs spatially close to each other are connected to a common air handling unit (AHU). A pressure sensor at the fans output provides an input to to a local control loop that regulates the electrical fan power to fix the fan pressure output. A network representation of a part of the physical system including conditioned spaces, fans and controllers is drawn in Figure~\ref{fig:Tag_network}; this data has been extracted from the tag data accompanying the recorded signals. This figure takes into account the spatial layout of conditioned rooms, and gives an idea of physical and communication links in the system.

Due to a conflict of local control loops, one of the fans (Fan 6 in Figure~\ref{fig:Tag_network}) in this building is behaving anomalously: at certain times of the day it produces mild uncontrolled oscillations. Although this action is not a result of a cyber attack, it represents a perfect initial test for the protocol aiming at detection and localization of failures: we expect that these oscillations should leave a signature in the correlations of related physical signals, while the signal is too weak to be visible and identified as an outlier in individual recorded signals. This anomalous behavior in the system is a proxy for attacks of the control architecture that can occur due to vulnerabilities of the cyber part of the network. First, we demonstrate the performance of our detection certificate, using the described Fan 6 oscillations as a failure event that we would like to detect and identify. At a second stage, we perform controlled experiments mimicking a simple intrusion on a smaller subset of devices in order to test the performance limits of the detection and localization algorithms as a function of the size of the anomalous set.

\begin{figure}[htb]
\centering
\includegraphics[width=1.0\columnwidth]{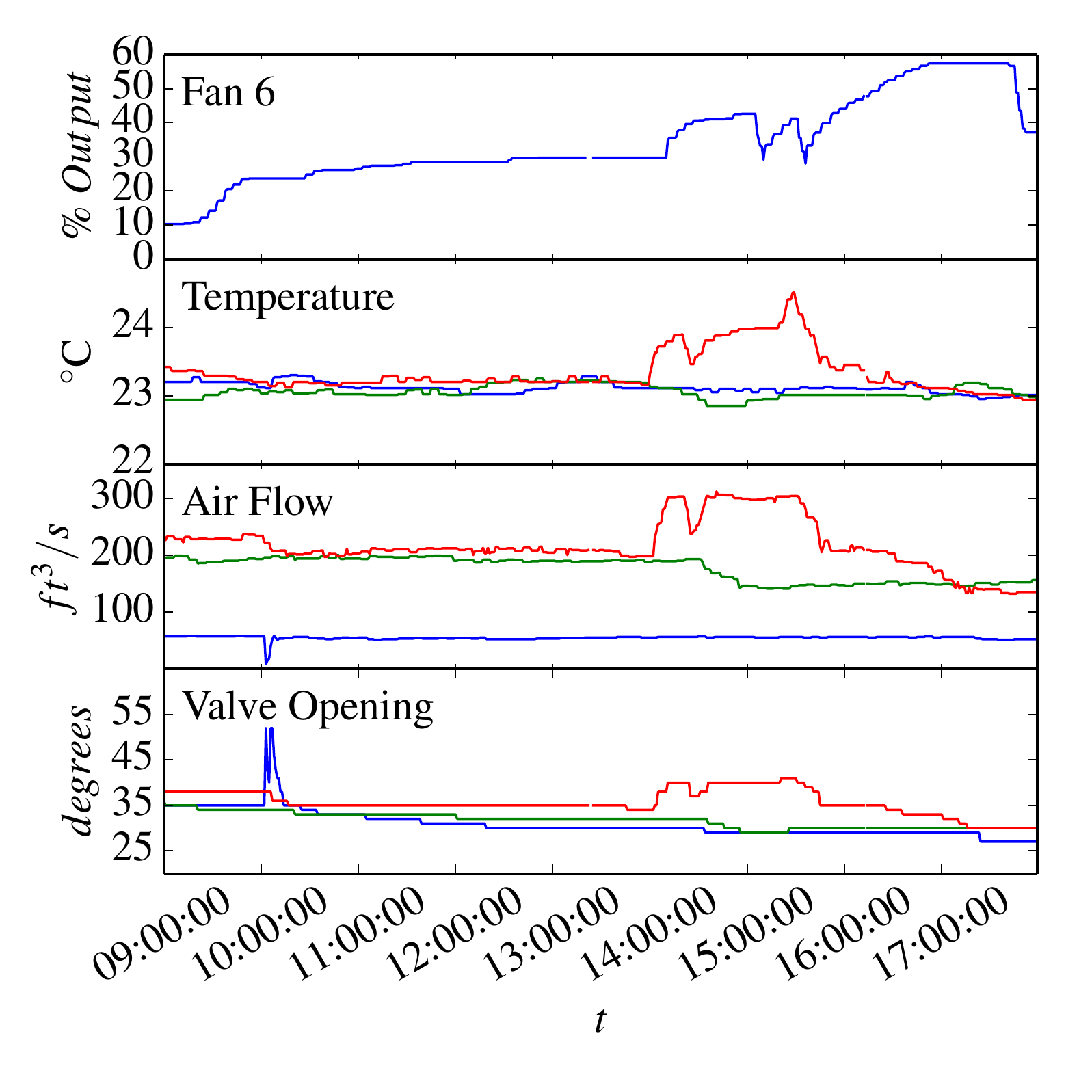}
\caption{Fan 6 oscillations create anomalous data measurements in rooms that are serviced by that fan.  Changes in output of Fan 6 can be seen in the temperature, air flow, and valve opening positions in Room 1 (red) and Room 2 (green) VAV measurement data but not in Room 3 (blue) data.  Room 1 and 2 are serviced by Fan 6 but Room 3 is not.
}
\label{fig:Signals}
\end{figure}

\subsection{Detection Algorithm Performance}

In Figure~\ref{fig:Signals}, we show examples of our data stream. The left plot of Figure~\ref{fig:Signals} shows an anomalous behavior of Fan 6, and three examples of temperature measurements in three conditioned spaces, two of which are serviced by Fan 6, and one being unrelated. The right plot shows examples of other signals of different types (airflow and valve positions) that we use for tests. The analysis of individual signals do not allow us to detect an anomalous behavior and to relate it to the malfunctioning Fan 6, and therefore we follow the procedure described in Section~\ref{sec:Time_Series}, constructing the correlation matrix and attempting to detect the anomaly from correlations of physical signals.

\begin{figure}[!ht]
\centering
\includegraphics[width=1.0\columnwidth]{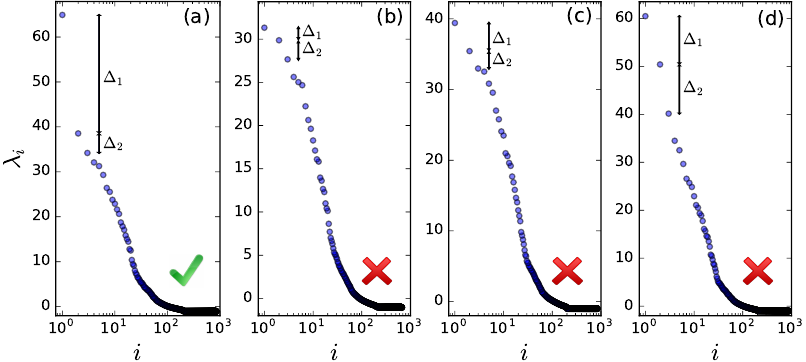}
\caption{Spectra (in the semi-log scale) of the correlation matrix $M$ for different scenarios. Oscillations of Fan 6 occur: (a) related signals included, (b) related signals excluded.  All signals included when (c) Fan 6 does not oscillate and (d) Fan 6 oscillates, but with smoothly with a large period.  Only the spectrum (a) satisfies the detection condition~\eqref{eq:Detection_condition}, as it should be.}
\label{fig:Detection}
\end{figure}

Let us first demonstrate the performance of the detection algorithm, described in Section \ref{sec:Detection}. In Figure \ref{fig:Detection}, we show the spectra of the correlation matrices $M$ in four different situations: i) Fan 6 oscillating, and all signals included; ii) Fan 6 oscillating, and signals serviced by Fan 6 removed from the data; iii) Fan 6 not oscillating, all signals included; iv) Fan 6 oscillating smoothly with a large period (on the order a half a day). It is clear that only case i) should trigger a positive detection outcome. Indeed, we notice that only the spectrum in this case satisfies the condition \eqref{eq:Detection_condition}, while all other situations yield a negative detection result. The matrix $M$ in each case has been constructed using the parameters $\tau_{\text{av}} = 30$ min and $\tau_{\text{corr}} = 200$ min.

\subsection{Localization Algorithm Performance}
Once the presence of anomaly is detected, we compare the performance of localization algorithms. Is it possible to correctly identify the group of nodes related to the anomalous fan, and hence to infer the reason of misbehavior? Tables~\ref{table:Tau_av_dependence} and~\ref{table:Tau_corr_dependence} demonstrate localization results for two values of group sizes. The ground truth $k_{0} = 209$, which is in general unknown, and for $k^{*} = 30$ strongest signals.  We follow the strategy outlined in Sections \ref{sec:Low-rank} and~\ref{sec:Biclustering} and use different combinations of the smoothing window time $\tau_{\text{av}}$ and the correlation time window $\tau_{\text{corr}}$. As discussed in Section~\ref{sec:Time_Series}, little relevant information is captured with small $\tau_{\text{av}}$, and indeed we find that $\tau_{\text{av}}=10$ does not lead to a positive detection, see Table~\ref{table:Tau_av_dependence}. The best results are obtained for larger values of $\tau_{\text{av}}$, where more data is incorporated in the correlation matrix.
\begin{table}[!ht]
\centering
\begin{tabular}{|c|c|c|c|c|c|c|c|} \hline
\multirow{3}{*}{$\tau_{\text{av}}$}&\multirow{3}{*}{Detection}&\multicolumn{6}{|c|}{Number of false positives} \\
\cline{3-8}
& & \multicolumn{2}{|c|}{\cellcolor{Gray} Low-rank}& \multicolumn{2}{|c|}{\cellcolor{Gray} $\mathcal{LAS}$}& \multicolumn{2}{|c|}{\cellcolor{Gray} $\mathcal{IGP}$} \\
\cline{3-8}
& & $k^{*}$ & $k_{0}$ & $k^{*}$ & $k_{0}$ & $k^{*}$ & $k_{0}$ \\ \hline
\rowcolor{Gray}
10 & \ding{55} & $27$ & $169$ & $26$ & $144$ & $25$ & $149$ \\ \hline
30 & \ding{51} & $0$ & $123$ & $0$ & $112$ & $0$ & $115$ \\ \hline
\rowcolor{Gray}
50 & \ding{51} & $0$ & $106$ & $0$ & $107$ & $0$ & $108$ \\
\hline\end{tabular}
\caption{Performance of different localization algorithms as a function of $\tau_{\text{av}}$ in the presence of Fan 6 activity. There are $k_{0}=209$  heterogeneous streams  serviced by Fan 6, out of $N=974$ total signals. The table demonstrates the number of mismatches (false detections) identified by the algorithms in the case of searched groups of sizes $k^{*}$ and $k_{0}$, with $k^{*} = 30$. For all cases, $\tau_{\text{corr}}=120$ min is kept fixed.}
\label{table:Tau_av_dependence}
\end{table}
\begin{table}[!ht]
\centering
\begin{tabular}{|c|c|c|c|c|c|c|c|} 
\hline
\multirow{3}{*}{$\tau_{\text{corr}}$}&\multirow{3}{*}{Detection}&\multicolumn{6}{|c|}{Number of false positives} \\
\cline{3-8}
& & \multicolumn{2}{|c|}{\cellcolor{Gray} Low-rank}& \multicolumn{2}{|c|}{\cellcolor{Gray} $\mathcal{LAS}$}& \multicolumn{2}{|c|}{\cellcolor{Gray} $\mathcal{IGP}$} \\
\cline{3-8}
& & $k^{*}$ & $k_{0}$ & $k^{*}$ & $k_{0}$ & $k^{*}$ & $k_{0}$ \\ \hline
\rowcolor{Gray}
90 & \ding{51} & $2$ & $128$ & $2$ & $120$ & $2$ & $122$ \\ \hline
120 & \ding{51} & $0$ & $123$  & $0$ & $112$ & $0$ & $115$ \\ \hline
\rowcolor{Gray}
160 & \ding{51} & $0$ & $112$  & $0$ & $110$ & $0$ & $109$ \\ \hline
200 & \ding{51} & $0$ & $106$  & $0$ & $103$ & $0$ & $104$\\
\hline\end{tabular}
\caption{Comparison of the localization algorithms under the same conditions as the ones described in Table~\ref{table:Tau_av_dependence}, as a function of $\tau_{\text{corr}}$. In this table, $\tau_{\text{av}}=30$ min is kept fixed.}
\label{table:Tau_corr_dependence}
\end{table}

One of the major requirements for the algorithms is the ability to perform online detection and localization. New data points arrive every minute, so we would like the localization algorithms to converge in several seconds. The low-rank algorithm is very fast, and does not need any adjustments. As discussed in the previous section, in order to meet the computation complexity requirement for the biclustering algorithm we are forced to limit the number of warm starts to $1000$ for the size $k_{0}=209$ and to $10000$ for $k^{*}=30$ since the convergence time of biclustering procedure grows with $k$. Another important property of the biclustering methods is that unlike in the low-rank approximation, the identities of the discovered columns do not always match the identity of the discovered rows; we use only one of the subsets to compute the number of mismatches.

With these restrictions, the three algorithms produce similar results with a comparable speed (under $3$ seconds for low-rank algorithm and within $20-30$ seconds for biclustering in the present case). While only half of the true nodes are discovered when searching for all of the $k_0$ anomalous signals, very few false positives occur when only searching for the $k^{*}$ strongest signals.  The discovered $k^{*}$ signals in almost all cases belong to a subgroup of a true group related to the anomalous fan. This value is sufficient to determine the common functional role of nodes inside this group, which corresponds to their relation to the anomalous Fan 6 in this case study. Therefore, all algorithms satisfy the requirements of performance, simplicity and scalability, which make them appropriate for deployment in real cyber-physical systems. In the next section, we discuss controlled experiments which would allow us to investigate the effect of the size of the anomalous community.

\subsection{Identification limits from controlled experiments}

Previously, we have tested the performance of the scheme on detecting the faulty behavior of Fan 6 already present in the system. In this section, we report results from controlled experiments on particular sensors of the office automation system. In their simplest form, these experiments consisted in a manipulation of temperature set points, mimicking localized intrusions of small amplitude. The trials were conducted on the controllers related to a small number of sensor units on Fan 5 (a non-oscillating fan, see Figure~\ref{fig:Tag_network}), while all sensors related to the anomalous Fan 6 have been excluded to avoid an undesired interference.

The experiments that we report here took the following form: the temperature set points for 10 chosen VAV were raised $0.5\degree F$ for $30$ minutes and then lowered $1\degree F$ for the next $30$ minutes. Each VAV contains three sensors measuring temperature, airflow, and valve opening position. The experimental intrusions potentially affected a total of $30$ data streams.
Among these $30$ data streams of interest, only $16$ showed a significant level of correlation. There are several reasons for this behavior, but the most important one consists in the observation that the airflow and valve opening positions have a much faster response to the set-point change compared to the temperature measurements which rise or fall on a much longer time scale. In the following we assume that these $k_{0} = 16$ sensors constitute the ground truth for an anomalous group of nodes.

Using the collected data, we validate the choice of $k^{*} = \sqrt{N}$ put forward in Sections~\ref{sec:Low-rank} and~\ref{sec:Biclustering}, and used throughout the study of the anomalous sensors related to the Fan 6. In particular, we verify that if the size of the group $k_{0}$ represents a sufficiently small fraction of the total number of signals, then it can not be correctly localized. In order to perform this study, we have considered $1000$ selections of $N$ randomly chosen signals but always containing the $k_{0}=16$ anomalous nodes. We applied our detection and localization protocol in each case for a range of $N$. The low-rank algorithm was used for localization as we have seen that at these scales it gives the same results with the fastest computation time; other localization methods show equivalent results. Note that the localization procedure was triggered only when the detection condition \eqref{eq:Detection_condition} was satisfied.

\begin{figure}[htb]
\centering
\includegraphics[width=1.0\columnwidth]{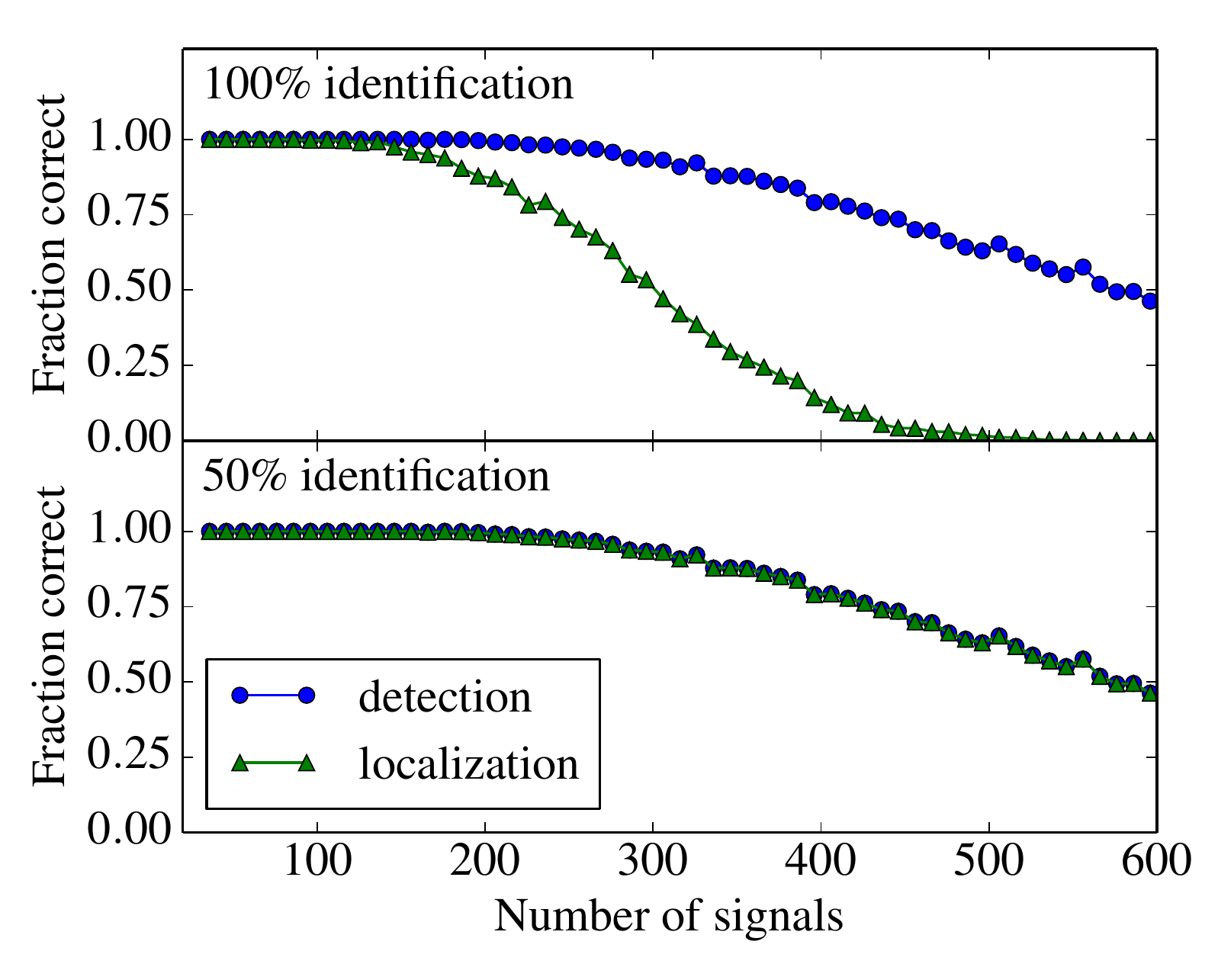}
\caption{Empirical probability of successful detection and localization of a group of $k_{0}=16$ anomalous devices as a function of the total number of signals $N$. Localization is considered as succesfull if all $k_{0}$ nodes are correctly identified (top) and if at least $50 \%$ of nodes are recovered (bottom). Each point is averaged over $1000$ random selections of $N$ signals.}
\label{fig:limits}
\end{figure}

The results are presented in the Figure~\ref{fig:limits} with the empirical probability of successful detection and localization shown as a function of the total number of signals $N$. Two definitions of success are examined;  a full and correct $100\%$ identification of the ground truth, and a successful localization of at least $50 \%$ of the $k_{0}$ nodes, i.e. correctly identifying at least $8$ devices out of $16$. For the $100\%$ identificaiton case we find a phase transition-type behavior as a function of $N$. The localization algorithm starts to fail at some point near $N = k_{0}^{2}$. This behavior is very close to the theoretical bounds derived in the idealized situations of Gaussian and Bernoulli distributions; in particular, it justifies our choice for $k^{*}$ in the case where the optimal community size is unknown. The second case of $50\%$ identificaiton  illustrates that if we allow for some mistakes in the identification of anomalous sensors, then a successful localization occurs every time the detection procedure yields a positive result. This procedure might be appropriate if the labeled network is sufficiently sparse and the common cause of the anomaly can be easily identified using the sensor labels even in the case where not all the nodes are correctly localized.

%

\section{Related Work}

{\bf Defense of cyber-physical systems:} Methods for detecting and localizing cyber-physical failures and attacks have attracted significant attention~\cite{sha2009cyber,shi2011survey,sharma2014modeling,mitchell2015modeling}.
Major hurdles stem from a high degree of influence of sensor data from seasonal changes, proximity correlations and operational switches, and from the fact that infrastructure operators do not always have an accurate model of the physical network (the assumption we make in this work), or the existing models are not integrated into unified cyber-physical system model \cite{sharma2014modeling}. Another important factor is an increasing size and complexity of the systems under considerations \cite{shafer2012rainmon}. Some of the previous works develop detection techniques based on an accurate system modeling and on accounting for different attack scenarios \cite{mitchell2015modeling}, which represents a completely different approach to the problem compared to the present study.

{\bf Signal detrending:} Aiming at general applications, we have used a simple running-mean signal detrending procedure in Section~\ref{sec:Time_Series}. The goal of detrending any time series $[X(t)]_{t=\tau}^{\tau+T}$ is to decompose the signal into a superposition of simpler pieces. There are a wide array of detrending methods \cite{brillinger2001time,box2015time,dickey1981likelihood,chatfield1978holt,gardner1985exponential,hyndman2008forecasting}
, and each have associated strengths and weaknesses. 
These detrending methods assume the time series is stationary which is most often achieved with a regression-line fit to the observed time series. 
After removing this trend the residual time series is evaluated for stationarity (i.e. $\mathbb{E}X(t) = \mathbb{E} X(t+\tau) \;; \tau \in \mathbb{N}$ ) using a Dickey-Fuller test \cite{brillinger2001time,box2015time,dickey1981likelihood}. A stationary signal can be further decomposed by assuming it follows a linear auto-regressive process \cite{brillinger2001time,box2015time}.  
An auto-regressive process is one that supposes the signal at time $t$ is a linear addition of the signal sampled at past time points $X(t) = \sum_{i=t-1} a_{i}X(i)$. 

Other data-driven approaches considered for detrending a times series are exponential-smoothing, for example, the Holt-Winters methodology \cite{chatfield1978holt}. Exponential and Holt-Winters smoothing detrend the time series by assuming the signal at time $t$ is made up of past observations weighted by a geometrically decreasing parameter $\alpha \in (0,1)$ such that $X(t) =  \alpha X(t-1) + (1-\alpha) s_{t-1}$ where $s_{t-1}$ is the cumulative sum of past weighted observations 
\cite{gardner1985exponential,hyndman2008forecasting}. 


{\bf Outliers detection:} Anomaly detection is an important field with application to a wide number of domains (see \cite{chandola2009anomaly} for a general survey). A large number of methods have been suggested, including network \cite{zhang2010outlier} and time series \cite{6684530} specific techniques. A general formulation of the anomaly detection problem often takes form of hypothesis testing by considering $H_{0}$ (absence of anomaly) versus $H_{1}$ (presence of anomaly). In the present work, the hypothesis $H_{1}$ has been formulated as follows: if the correlation matrix is constructed and normalized in such a way that the normally behaving correlations fluctuate around zero, then there exist a submatrix with elements having a deviating mean \cite{ma2015computational}. This task is directly related to the problem of finding hidden cliques and community detection in graphs \cite{Fortunato201075}.

{\bf Optimal denoising:} Real-world correlation matrices are noisy, and in general it is not sufficient to work directly with the observed data. One should develop techniques for extracting a useful signal from the signal-plus-noise matrix, the procedure also known as denoising which appears in many machine learning \cite{Kannan2009}, signal processing \cite{scharf1991svd} and classification applications \cite{klema1980singular}. Moreover, in reality the signal matrix might have no special  structure, while the form of the noise term is in general unknown. Several studies have explored the problem of the effective rank estimation of the signal matrix by optimal thresholding of singular values \cite{nadakuditi2014optshrink, chatterjee2015matrix}. In this work, we encountered a different problem of estimating the size of the anomalous submatrix under the rank 1 assumption.

\section{Conclusions}

In this work we explored a set of methods for detection and localization of failures in cyber-physical systems, based on the analysis of correlations between physical time series. The established protocol enables the identification of a group of anomalous sensors and provides insight for the localization of the failure source. The developed detection procedure achieves a number of important requirements, including low computational complexity and simplicity of implementation. Our capability to access the cyber-physical demonstration system, described in the article, to collect and analyze data from this system, and to deploy the presented detection algorithm opens a path forward for future work. We plan to continue real-world experiments which will consist of manipulating the building control system in a known manner using diverse attack strategies; this will allow us to further validate the presented methods. Another direction that we intend to explore consists of combining more control communication network data in order to minimize the possibility of false detections and to enhance the quality of failure source localization. These developments are essential for conception of algorithms for proportional response and for designing resilient cyber-physical networks.


\section{Acknowledgments}

The authors acknowledge Arthur Barnes, Gary Goddard and Hari Khalsa for their help with data collection, and Charles Bordenave, Michael Chertkov, David Gamarnik, Earl Lawrence, Sidhant Misra and N. Raj Rao for fruitful discussions. This work was funded by the Department of Energy at Los Alamos National Laboratory under contract DE-AC52-06NA25396 through the Laboratory-Directed Research and Development Program.

\bibliographystyle{abbrv}
\bibliography{cyberphysical}
\end{document}